\documentclass[%
reprint,
nofootinbib,
 amsmath,amssymb,
 aps,
]{revtex4-2}

\usepackage{graphicx}
\usepackage{dcolumn}
\usepackage{bm}

\begin{document}

\preprint{APS/123-QED}

\title{Nonreciprocal metasurfaces with epsilon-near-zero materials}


\author{Albert Mathew$^1$, Rebecca Aschwanden$^2$, Aditya Tripathi$^1$, Piyush Jangid$^1$, Basudeb Sain$^{2,3}$, Thomas Zentgraf$^{*2}$, Sergey Kruk$^{*4}$}
\affiliation{$^1$Research School of Physics, The Australian National University,
Canberra 2601, Australia}

\affiliation{$^2$Department of Physics, Paderborn University, Paderborn 33098, Germany}
\affiliation{$^3$Department of Physics and Electronics, Christ University, Bangalore, 560029, India}

\affiliation{$^4$Australian Research Council Centre of Excellence QUBIC, IBMD, School of Mathematical and Physical Sciences, University of Technology Sydney 2007, Australia}
\email{thomas.zentgraf@uni-paderborn.de}
\email{sergey.kruk@outlook.com}

\date{\today}

\begin{abstract}
Nonreciprocal optics enables asymmetric transmission of light when its sources and detectors are exchanged. A canonical example -- optical isolator -- enables light propagation in only one direction, similar to how electrical diodes enable unidirectional flow of electric current. Nonreciprocal optics today, unlike nonreciprocal electronics, remains bulky. Recently, nonlinear metasurfaces opened up a pathway to strong optical nonreciprocity at the nanoscale. However, demonstrations to date were based on optically slow nonlinearities involving thermal effects or phase transition materials. In this work, we demonstrate a nonreciprocal metasurface with an ultra-fast optical response based on indium tin oxide in its epsilon-near-zero regime. It operates in the spectral range of 1200-1300 nm with incident power densities of 40-70 GW/cm$^2$. 
Furthermore, the nonreciprocity of the metasurface extends to both amplitude and phase of the forward/backward transmission opening a pathway to nonreciprocal wavefront control at the nanoscale.

\end{abstract}
\maketitle

Contemporary optics is undergoing revolutionary transformations driven by nanotechnology. Today we can nanofabricate functional optical components hundreds of times thinner than a human hair that outperform conventional bulky optics \cite{Kruk2017,Kamali2018AControl,Chen2020FlatMetasurfaces}. A vital but largely unaddressed problem of contemporary nanoscale optics and nanophotonics is how to break reciprocity\cite{Peng2014WhatMicrocavities,Caloz2018ElectromagneticNonreciprocity} in light propagation. Nonreciprocity is instrumental for the traffic control of signals in high-end laser systems, optical communications and machine vision with LiDAR technology \cite{Yang2020Inverse-designedLiDAR}.

However, optical nonreciprocity is difficult to achieve and there exist only three known conceptual pathways \cite{Peng2014WhatMicrocavities,Caloz2018ElectromagneticNonreciprocity}: (i) materials with asymmetric permittivity/permeability tensors such as ferrites, (ii) time-varying systems, and (iii) nonlinear light-matter interactions. 

Today virtually all commercial optical isolators are based on the first pathway of materials with asymmetric material property tensors that show a magneto-optic Faraday effect under a static magnetic field. Such optical isolators require strong magnets, which hinder their miniaturization. Typical optical isolators today measure centimetres in size. Materials with high remanent magnetization (magnetism that remains in a material after an external magnetic field is removed) may represent a pathway toward miniaturization of nonreciprocal optics. Remanent magnetisation was used in nonreciprocal metasurfaces in the microwave spectral range (at around 15GHz frequency or 2 cm wavelength) \cite{Yang2023}. This is, however, challenging to apply at optical frequencies due to constrains of material properties. As of now, miniaturization of nonreciprocal optical components based on materials with asymmetric permittivity/permeability tensor faces a roadblock. This prevents fabrication and integration of large quantities of nonreciprocal photonic components into intricate layouts similar to how nonreciprocal electronic components -- diodes and transistors -- are integrated into chips.

The second approach based on time varying systems \cite{Yu2009CompleteTransitions,Kang2011ReconfigurableFibre,Estep2014Magnetic-freeLoops} includes a broad range of phenomena covering acousto-optics \cite{Kittlaus2020,Tian2021}, opto-mechanics \cite{Wanjura2023}, and electro-optics \cite{Yu2023}. The advancements in time-varying systems led to a significant miniaturization of nonreciprocal optics down to the micrometre-scale. Recent demonstrations include systems based on lithium niobate \cite{Sohn2021ElectricallySplitting} and silicon nitride on-chip waveguiding platforms \cite{Ji2022CompactCircuits}. However, further miniaturisation of such systems from the micro-scale to the nanoscale has not been achieved.

At  present, the most feasible pathway towards nonreciprocity at the nanoscale is via nonlinear light-matter interactions. Nonlinearity-induced nonreciprocity at the sub-micrometer scale has been studied in unstructured thin films \cite{Tocci1995ThinfilmDiode,Anand2013OpticalDevice,Tang2020BroadbandDiode,Wan2018LimitingDioxide}. However, such observations were accompanied by low levels of transmission and high insertion losses, which hindered their development beyond initial proof-of-concept experiments. Nonlinearity-induced nonreciprocity began rapidly evolving at its confluence with the physics of nanoresonators and metasurfaces. Optical metasurfaces can be realised as arrays of subwavelength elements judiciously designed to resonate with the incident light thus enhancing light-matter interactions by orders of magnitude \cite{Koshelev2020SubwavelengthNanophotonics, Zubyuk2021ResonantFields}. 
Theoretical studies of nonlinear nanoresonators and metasurfaces revealed asymmetric and nonreciprocal responses of light \cite{Poutrina2016MultipolarGeneration,Kim2021AsymmetricMetasurface,Lawrence2018NonreciprocalMetasurfaces,Jin2020Self-InducedMetasurfaces,Cheng2020SuperscatteringAntennas,Antonellis2019NonreciprocityComponents,Karakurt2010NonreciprocalSurfaces}. Experimental demonstrations of nonlinear effects that are asymmetric with respect to forward vs backward propagation of light were performed for the generation of optical harmonics of third \cite{Shitrit2018AsymmetricMetasurfaces,Kruk2022AsymmetricMetasurfacesb,Kruk2019NonlinearNanostructures} and second \cite{Boroviks2023DemonstrationPseudodiode} order. Nonlinear self-action leading to nonreciprocity was recently demonstrated in metasurfaces hybridised with a phase-change material (vanadium dioxide) in both near-infrared  spectral regions (fibre communications band) \cite{Tripathi2024NanoscaleMetasurfaces} and in
mid-infrared \cite{King2024ElectricallyIsolation}, as well as in silicon grating metasurfaces leveraging a combination of thermal and Kerr-type effects \cite{Cotrufo2024PassiveContinuum}. Such nonlinear metasurfaces, being fundamentally self-biased systems (that is, the nonreciprocal response being triggered by the propagating light itself without any external stimuli), have a finite switching time between different responses in "forward" and "backward" directions. In ref.\cite{Cotrufo2024PassiveContinuum} switching times on the order of microseconds were reported. The switching time in the vanadium dioxide metasurfaces was theoretically estimated \cite{Tripathi2024NanoscaleMetasurfaces} to be in the range of microseconds for the continuous wave operation going down for pulsed operation to sub-nanoseconds (transmission fall time) and hundreds of nanoseconds (transmission rise time). To date, nanoscale nonreciprocity was not demonstrated in a platform enabling ultra-fast optical switching times.

Here, we demonstrate an optical metasurface that incorporates an epsilon-near-zero (ENZ) material\cite{alam2016large}. Such ENZ regimes of optical responses were demonstrated in multiple material platforms including indium tin oxide (ITO) \cite{alam2016large} and aluminium zinc oxide \cite{Caspani2016EnhancedMaterials}. ENZ materials incorporated into metasurfaces were shown to enhance light confinement and to boost nonlinear effects\cite{shi2021polarization,deng2020giant, xie2024strong,minerbi2022role,alam2018large, ma2023active,schulz2020optical}, enabling optical modulation\cite{vatani2022all}, broadband optical absorption\cite{wang2019electrically}, optical switching\cite{xie2020tunable}, photon acceleration\cite{liu2021photon}, and electrochemical control\cite{kaissner2021electrochemically,karst2021electrically}. We note that the nonlinear response of epsilon-near-zero materials can be on the femtosecond scale. In our work, the nonlinear refractive index modulation of the ENZ material leads to ultra-fast nonlinear nonreciprocity. We choose ITO as the material platform and fabricate silicon (Si) nanodisks on top of it such that the combined system of the Si-nanodisks and the ITO-film forms a nano-resonator (Figs. 1a-c). 
The measured relative permittivity of the ITO film shows the transition from positive to negative values at a wavelength of around 1230 nm (Fig. 1d).
The asymmetry of the design in the direction perpendicular to the metasurface (the direction of the light propagation) leads to difference in the near-field distributions within the resonator for "forward" and "backward" directions (see Fig. 2). In the regime of linear optics this results in identical far-field transmission in forward and backward direction as demanded by reciprocity. However, when nonlinear light-matter interactions are factored, different near-field distributions result in different levels of modulation of the refractive index of the ENZ material for "forward" and "backward" scenarios. This, in turn, results in different transmission levels leading to nonreciprocity in both amplitude and phase of the transmission.

\begin{figure*}[bth!]
 \centering
 \includegraphics[width=0.5\textwidth]{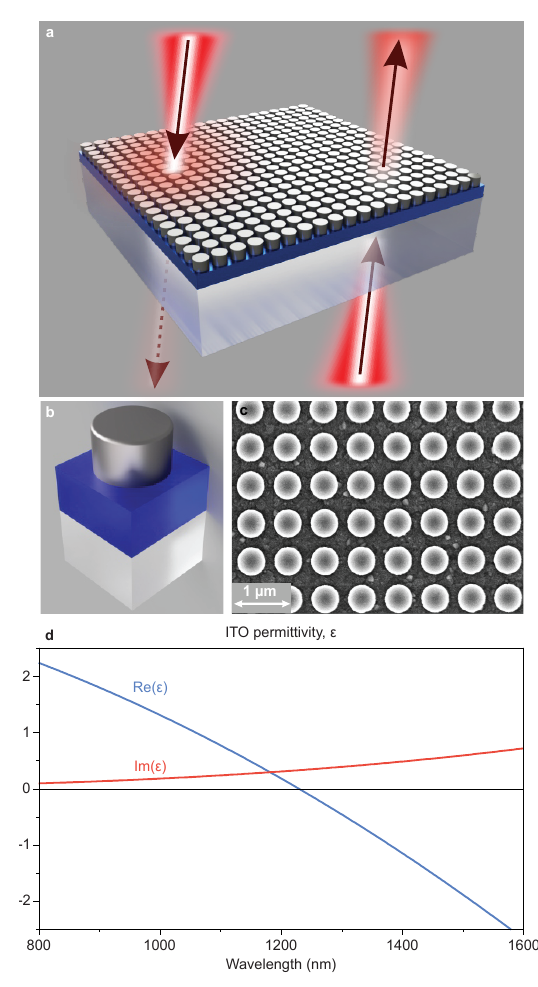}
 \caption{Nonreciprocal transmission of light
through a hybrid Si-ITO metasurface. (a) Concept
image of nonreciprocal transmission through a metasurface.
(b) Schematics of a subwavelength resonator
(metasurface unit cell): silicon disk placed on top of ITO and glass film. 
(c) Electron microscope image of the fabricated Si-ITO metasurface. (d) Experimentally measured optical constants of the ITO.}
 \label{figure_1}
\end{figure*}



\begin{figure*}
 \centering
 \includegraphics[width=0.49\textwidth]{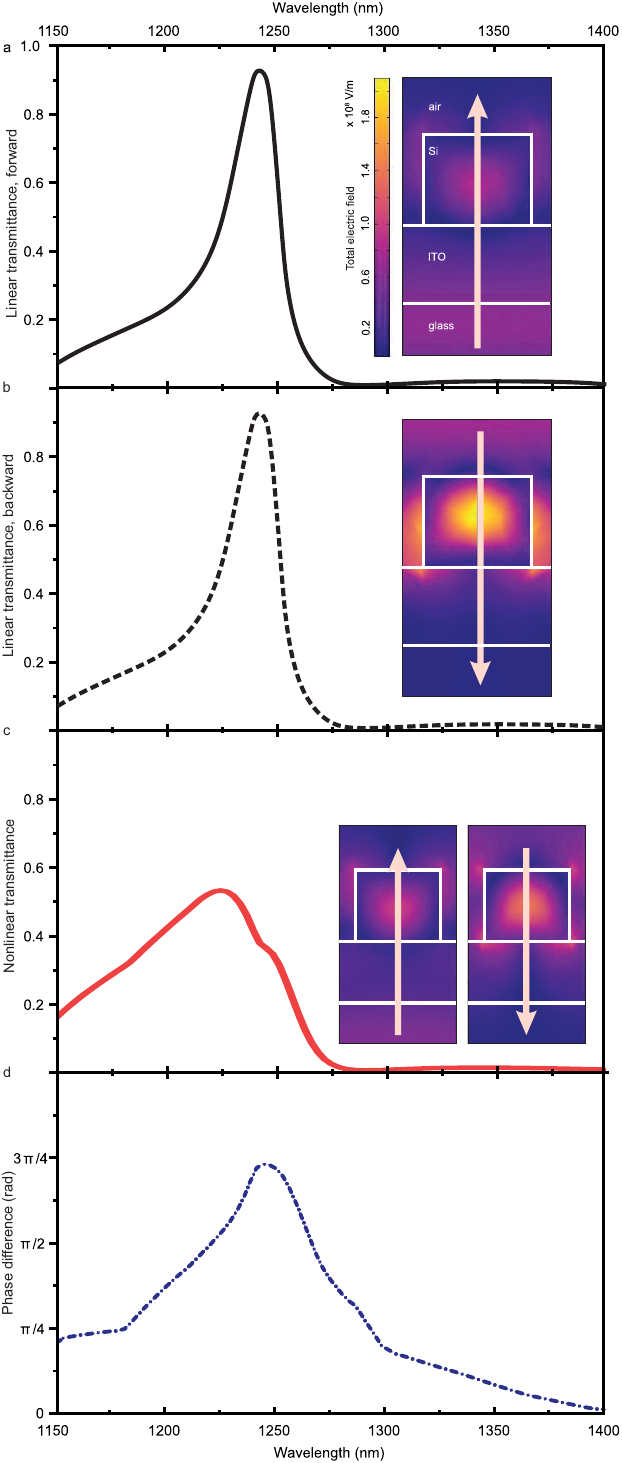}
 \caption{Numerical results for Si-ITO metasurface. 
 (a) Calculated linear forward transmission. Inset: corresponding near-field distribution of the amplitude of the electric field at 1230 nm wavelength shown over the cross-section of the metasurface unit cell. Arrow visualizes the direction of propagation.
 (b) Calculated linear backward transmission. Inset: corresponding near-field at the 1230 nm wavelength with the arrow visualizing the direction of propagation. (c) Calculated nonlinear transmission assuming refractive index change of the ITO in the vicinity of the ENZ region. The nonlinear portion of the refractive index was taken from ref. \cite{alam2016large} for 50 GW/cm$^2$ peak power density. Line show identical forward/backward transmission. Insets show the corresponding near-fields. (d) Calculated contrast of phase accumulation in transmission between the linear and nonlinear regimes of the ITO film.}
 \label{figure_2}
\end{figure*}

We optimise this effect with the design of the metasurface in COMSOL by maximizing simultaneously two parameters: the metasurface transmission in the linear regime and the difference in the field concentration inside the ITO film for the "forward" and "backward" scenarios at the ENZ wavelength of the ITO material (in our case 1230 nm). The linear optical constants (real and imaginary parts of the refractive index) for ITO used in the simulation are measured experimentally\cite{mathew_2024_14208280}. The refractive index value for silicon\cite{pierce1972electronic} and the fused silica substrate\cite{malitson1965interspecimen}, are obtained from the literature. The resulting optimised design features a Si disk 357 nm in height and 458 nm in diameter arranged into a square lattice with 636 nm period and residing on an ITO film with 310 nm thickness. The sample rests on a silicon oxide glass substrate that is considered as semi-infinite in the simulations.

Figure 2a shows the linear transmission (at low light intensities) through the metasurface in forward direction. Figure 2b shows the identical linear transmission in backward direction as expected from reciprocity. The insets in the Figs. 2a,b show the corresponding near-field distributions of the amplitude of the electric field for both cases at 1230 nm wavelengths. We notice that although, the far-field transmission is identical for both forward and backward directions, the corresponding near-fields are drastically different. Once nonlinear interactions are factored, the drastic difference in the near-field distributions becomes the key component in achieving different and nonreciprocal modulation of the ITO refractive index for "forward" vs. "backward" propagation.

Figure 2c shows the transmission through the metasurface in the nonlinear case when the ITO refractive index is modulated by 50 GW/cm$^2$ peak power density with the nonlinear portion of the refractive index taken from \cite{alam2016large}. The modulation of the refractive index of the ITO film translates into a pronounced modulation of the transmission amplitude. 

Additionally, we calculated the phase difference between the linear and nonlinear regime of the transmitted light (Fig. 2d). We observe the largest difference in phase accumulation at around ENZ wavelength of 1230 nm where the nonlinear modulation of the ITO refractive index is the strongest.

We fabricate the designed metasurface on top of a commercial ITO film. Silicon is deposited by plasma-enhanced chemical vapour deposition and patterned by electron beam lithography. The lithography resist negative mask is converted into metal positive mask (chromium) using thermal evaporation and lift-off. Reactive ion etching is used to translate the mask pattern into the Si film to form the Si disks. Finally, the metal mask is removed by selective etching. An electron microscope image of the resulting sample is shown in Fig. 1c.

\begin{figure*}[bth!]
 \centering
 \includegraphics[width=1.0\textwidth]{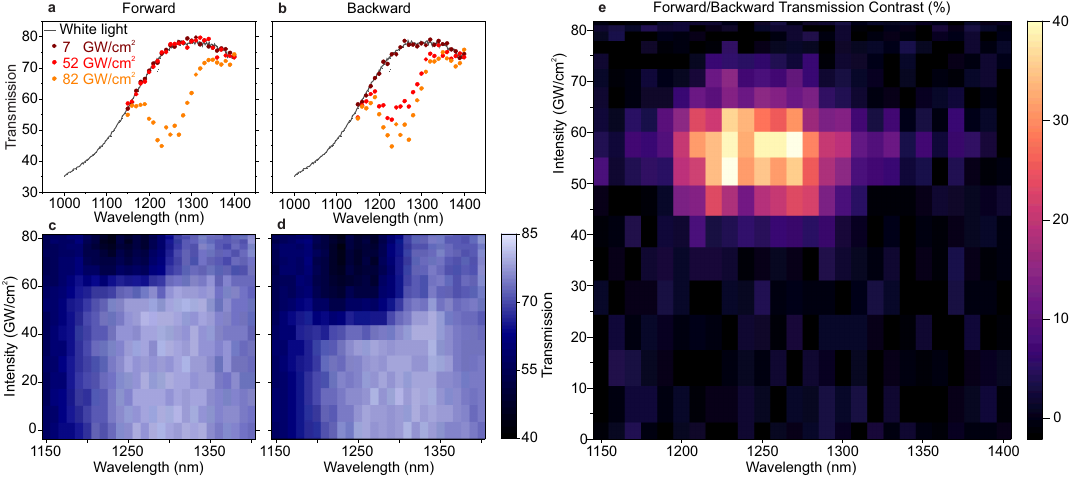}
 \caption{Experimental results. (a,b) White light transmission and nonlinear transmission at incident power of 7, 52 and 82 GW/cm$^2$ for (a) forward (b) backward illuminations.  (c,d) Wavelength and power dependence of the nonlinear transmission for (c) forward and (d) backward illuminations. (e) Wavelength and power dependence of forward to backward transmission ratio showing the highest contrast around 1200-1300 nm wavelengths and intensities around 40-70 GW/cm$^2$.}
 \label{figure_3}
\end{figure*}

In the next step, we pre-characterise the metasurface transmission using a thermal light source and an Ocean Optics QE-Pro spectrometer. The resulting white-light spectra (identical for "forward" and "backward" directions as expected from reciprocity) are shown in Figs. 3a,b as the black curves. To switch between the forward/backward experiments we flip the sample inside the setup. Next, we illuminate the metasurface with short laser pulses from an optical parametric amplifier (Coherent Opera) with pulse duration of 230 fs, repetition rate of 1 MHz and tunable wavelength. A small portion of the laser beam is reflected onto a power meter tracking the incident intensity. The power is attenuated with a set of a waveplate and a polariser. The laser beam is then focused onto the metasurface with a Mitutoyo PlanAPO 100$\times$ NIR objective. The transmitted signal is captured by an identical objective. We measure the intensity of light transmitted through the metasurface with a second power meter and reference it to the transmission through the ITO-coated glass substrate. Figures 3a,b show the transmission through the metasurface for forward and backward scenarios of excitation at three different incident power levels. We observe that for the peak power densities of 7 GW/cm$^2$ (dark red points in Fig. 3a,b), both "forward" and "backward" transmissions remain identical to each other and to the linear white-light transmission (black line). At this power level, the transmission is reciprocal as the nonlinear response of the ITO is insignificant. At higher power densities of 82 GW/cm$^2$ (orange dots in Figs 3a,b) the "forward" and "backward" transmissions are also similar to each other, however both notably lower than the linear transmission. At this power level, transmission is reciprocal as the nonlinear response of the ITO becomes saturated and thus similar for both "forward" and "backward" directions. At the intermediate power level of 52 GW/cm$^2$ the metasurface demonstrates a drastically different transmission for "forward" and "backward" excitations. In the "forward" direction, the transmission remains similar to the linear regime (black curve). In the "backward" direction the transmission deviates from the linear case significantly due to nonlinear modulation of the refractive index of the ITO. This manifests optical nonreciprocity arising due to different near-field distributions of the incident light inside the ITO film for "forward" and "backward" directions as predicted theoretically in Fig. 2c. 

A complete scan of the  forward/backward transmission behaviour in dependence of the incident peak power is shown  in  Figures 3c,d. From there one can see that the low transmission range around 1230 nm appears for lower peak power values for the "backward" transmission compared to the "forward" direction. 

Figure 3e shows the transmission contrast between "forward" and "backward" propagation as the function of wavelength and incident peak power. Pronounced nonreciprocal transmission occurs within the range of power densities of about 40-70 GW/cm$^2$ and within the spectral range of about 1200-1300 nm. Importantly, the spectral range of the nonreciprocal transmission is in the immediate vicinity of the ENZ wavelength of 1230 nm. For the region outside of the 1200-1300 nm range, the transmission becomes reciprocal and near-identical to the linear transmission, therefore the effect can be attributed only to the nonlinear ENZ response of the ITO film.


In summary, we demonstrated a new pathway to optical nonrecirocity at the nanoscale with ENZ-based metasurfaces. In our experiments, a Si metasurface on an ITO thin film operating in the ENZ regime at 1230 nm wavelength was used. Nonreciprocal optical transmission was measured within the spectral range of 1200-1300 nm and a peak power range of 40-70 GW/cm$^2$. The ITO material nonlinear response was previously estimated to be no longer than 200 fs (refractive index rise time) and 360 fs (recovery time) \cite{alam2016large}. This is consistent with the response time of the metasurface reported here operating under 230 fs pulsed excitation making the switching process ultrafast.

Single largest constrain of the current ITO-based nonreciprocal metasurface is the requirement for high peak power of light limiting the range of foreseeable applications to pulsed laser systems. This requirement is, however, of technical rather than fundamental nature. We note that nonlinearity-based nonreciprocity in metasurfaces can be achieved at much lower light intensities, e.g. down to hundreds of Watts per cm$^2$ (comparable to power density of a focused laser pointer) in phase-change materials\cite{Tripathi2024NanoscaleMetasurfaces}, albeit at a cost of slower response times. We foresee the advancements in all key parameters of nonreciprocal metasurfaces, including isolation values, response times and requirements on excitation powers brought by other types of epsilon-near-zero materials, such as PEDOT\cite{Karst2021ElectricallyNanoantennas} and PANI\cite{PANI} among others, as well as by judicious engineering of optical resonances within the metasurfaces, such as resonances associated with the bound states in the continuum\cite{Koshelev2020SubwavelengthNanophotonics}.

Importantly, in this work our theoretical analysis revealed a nonreciprocal phase accumulation within the metasurface. The ability to control the phase with metasurfaces and to create phase gradients  arguably enables some of the most significant advantages in meta-optics. Metasurfaces controlling phase of light evolved into functional flat optical components \cite{Kruk2017} such as lenses \cite{Capasso,paniagua2017metalens}, holograms \cite{zheng2015metasurface,wang2016grayscale} and beam deflectors \cite{yu2015high}. Nonreciprocal phase accumulation within metasurfaces opens up possibilities for a conceptually new class of optical components combining pairs of different and independent functionalities for the "forward" vs "backward" directions of light propagation. The optical nonreciprocity originates from the response of a single resonant unit cell of a subwavelength volume, which  opens up the design freedom for nonreciprocal metasurfaces assembled from dissimilar nanoresonators. Such design freedom gives access to spatially inhomogeneous and gradient nonreciprocal control over both amplitude and phase, similar to the complex asymmetric control of light achieved for parametric light generation \cite{Kruk2022AsymmetricMetasurfacesb}. Bias-free nanoscale nonreciprocal optics controlling forward-to-backward ratios of both amplitude and phase will facilitate the advancements of machine vision and optical signals routing and switching alongside with other technologies.

\section{Acknowledgments}
The authors acknowledge a financial support from the Australian Research Council (grant DE210100679) and the Australia-Germany Joint Research Cooperation Scheme (grant 57692192). T.Z. and R.A. acknowledge funding by the Deutsche Forschungsgemeinschaft (DFG, German Research Foundation) – TRR142/3-2022 -- No. 231447078 -- projects B09/A08.





\bibliography{main}

\end{document}